\documentclass[a4paper,12pt]{article}
\pdfoutput=1
\usepackage{geometry}
\usepackage{amsmath,amssymb,amsfonts}
\usepackage{xcolor,graphicx,cite,soul}
\usepackage{array,booktabs,longtable}
\usepackage{caption,microtype}
\usepackage{subcaption}
\usepackage{hyperref}
\usepackage{titlesec}
\usepackage{float}

\titleformat{\section}
  {\normalfont\fontsize{15}{12}\bfseries}{\thesection}{1em}{}

\hypersetup{
  colorlinks   = true, 
  urlcolor     = blue, 
  linkcolor    = black, 
  citecolor   = blue 
}
\usepackage[numbers,sort&compress]{natbib}
\let\OLDthebibliography\thebibliography
\renewcommand\thebibliography[1]{
  \OLDthebibliography{#1}
  \setlength{\parskip}{4pt}
  \setlength{\itemsep}{0pt plus 0.3ex}
}

\allowdisplaybreaks
\usepackage{tabularx}
\usepackage{bbm}
\usepackage{bm}
\usepackage{xfrac}
\usepackage{listings}
\usepackage{longtable}
\usepackage{datetime}
\usepackage{appendix}
\usepackage{ulem}
\usepackage{soul}
\usepackage{multirow}
\usepackage{cancel}
\usepackage{multicol}
\raggedcolumns
\makeatletter
\providecommand{\tabularnewline}{\\}

\makeatother

\newcommand{\gsim}
{\;\raisebox{-.3em}{$\stackrel{\displaystyle >}{\sim}$}\;}

\newcommand\al{\alpha}
\newcommand\be{\beta}
\newcommand\tb{\tan\beta}
\newcommand\TB{t_\beta}

\newcommand\CA{c_\alpha}
\newcommand\CBA{c_{\beta - \alpha}}
\newcommand\SBA{s_{\beta - \alpha}}

\newcommand\ReDiag{\mathop{%
  \raise .5pt\hbox{[}%
  \widetilde{\mathrm{Re}}%
  \raise .5pt\hbox{]}}}
\newcommand\ReOffDiag{\mathop{%
  \raise .5pt\hbox{$\llbracket$}%
  \widetilde{\mathrm{Re}}%
  \raise .5pt\hbox{$\rrbracket$}}}

\newcommand\SM{\ensuremath{\mathrm{SM}}}
\newcommand\THDM{\ensuremath{\mathrm{2HDM}}}

\newcommand\Mh{m_h}
\newcommand\MH{m_H}
\newcommand\MA{m_A}
\newcommand\MHp{m_{H^\pm}}
\newcommand\mbar{\bar{m}^2}

\newcommand\msq{m_{12}^{2}}

\newcommand\refeq[1]{Eq.~(\ref{#1})}

\newcommand\refta[1]{Tab.~\ref{#1}}
\newcommand\refse[1]{Sect.~\ref{#1}}

\newcommand\citere[1]{Ref.~\cite{#1}}
\newcommand\citeres[1]{Refs.~\cite{#1}}

\newcommand{\CP}{{\cal CP}}
\newcommand{\cp}{{\CP}}

\newcommand{\gev}{\,\, \mathrm{GeV}}

\newcommand\fb{\ensuremath{\,\mbox{fb}}}

\newcommand{\sig}{\sigma}

\def\reffi#1{\mbox{Fig.~\ref{#1}}}
\def\reffis#1{\mbox{Figs.~\ref{#1}}}

\def\la{\lambda}
\newcommand\kala{\ensuremath{\kappa_{\lambda}}}
\newcommand\laSM{\ensuremath{\lambda_{\mathrm{SM}}}}
\newcommand{\lahhh}{\ensuremath{\la_{hhh}}}
\newcommand{\lahhH}{\ensuremath{\la_{hhH}}}

\newcommand{\tree}{\ensuremath{^{(0)}}}
\newcommand{\one}{\ensuremath{^{(1)}}}

\newcommand{\mhh}{\ensuremath{m_{hh}}}

\definecolor{Orange}{named}{orange}
\definecolor{Purple}{named}{purple}
\definecolor{Lightblue}{cmyk}{0.9,0.1,0.1,0.3}
\definecolor{dgelborange}{cmyk}{0.,0.3,0.5, 0.}
\definecolor{Lila}{rgb}{0.5,0.,1}
\definecolor{Darkgreen}{rgb}{0.,.7,0.2}

\newcommand{\GW}[1]{{\color{black}#1}}

\newcommand{\KR}[1]{{\color{black}#1}}

\graphicspath{{figs/}}
\allowdisplaybreaks
\begin{document}
\thispagestyle{empty}

\def\thefootnote{\fnsymbol{footnote}}

\begin{flushright}
\mbox{}
DESY-24-041\\
IFT--UAM/CSIC-24-041 \\
KA-TP-04-2024 
\end{flushright}


\begin{center}

{\large\sc 
{\bf Higgs Pair Production in the 2HDM:
Impact of Loop Corrections\\[.5em] to the Trilinear Higgs Couplings and Interference Effects on\\[.5em] 
Experimental Limits}
}
\vspace{1cm}

{\sc
S.~Heinemeyer$^{1}$%
\footnote{email: Sven.Heinemeyer@cern.ch}%
, M.~M\"uhlleitner$^{2}$%
\footnote{email: margarete.muehlleitner@kit.edu}%
, K.~Radchenko$^{3}$%
\footnote{email: kateryna.radchenko@desy.de}%
~and G.~Weiglein$^{3,4}$%
\footnote{email: georg.weiglein@desy.de}%
}

\vspace*{.7cm}

{\sl
$^1$Instituto de F\'isica Te\'orica (UAM/CSIC), 
Universidad Aut\'onoma de Madrid, \\ 
Cantoblanco, 28049, Madrid, Spain

\vspace*{0.1cm}

$^2$Institute for Theoretical Physics,
Karlsruhe Institute of Technology, 76128 Karlsruhe, Germany

\vspace{0.1em}

$^3$Deutsches Elektronen-Synchrotron DESY, Notkestr.\ 85, 22607 Hamburg,
Germany

\vspace{0.1em}

$^4$II.\  Institut f\"ur  Theoretische  Physik, Universit\"at  Hamburg, Luruper Chaussee 149,\\ 22761 Hamburg, Germany
 
}

\end{center}

\vspace*{0.1cm}

\begin{abstract}
\noindent
The results obtained at the LHC for constraining
the trilinear Higgs self-coupling
of the detected Higgs boson at about 125 GeV, \lahhh, 
via the Higgs pair production process have 
significantly improved during the last years. 
We investigate the impact of 
potentially large higher-order corrections and interference effects on the comparison between the experimental 
results and the theoretical predictions for the pair production of the 125~GeV Higgs boson at the LHC.
We use
the theoretical framework of the Two Higgs Doublet Model (2HDM), 
containing besides the SM-like $\cp$-even Higgs boson $h$ a second $\cp$-even Higgs boson $H$, 
which we assume to be heavier, $\MH > \Mh$.
We analyze in particular the invariant mass distribution of the two produced Higgs bosons and show that
the loop corrections to the trilinear Higgs
couplings \lahhh\ and \lahhH\ as well as 
interference contributions give rise to important effects both for the differential and the total cross section.
We point out the implications for the 
experimental limits that can be obtained in the 2HDM for the case of the 
resonant production of the heavy Higgs boson $H$.
We emphasize the importance of the inclusion of interference effects between resonant and non-resonant
contributions in the experimental analysis for a reliable determination of exclusion bounds for a 
heavy resonance 
of an extended Higgs sector.

\end{abstract}

\def\thefootnote{\arabic{footnote}}
\setcounter{page}{0}
\setcounter{footnote}{0}


\section{Introduction}
\label{sec:intro}

After the discovery of a new scalar particle with a mass of 
about 125 GeV by ATLAS and CMS in 
2012~\cite{Aad:2012tfa,Chatrchyan:2012xdj,Khachatryan:2016vau}, 
several of its properties have meanwhile been measured with a remarkable precision.
From the results in particular for its couplings to the 
third generation fermions and to the massive gauge bosons 
it can be inferred that within the present experimental and 
theoretical uncertainties the predictions for the Higgs sector of 
the Standard Model (SM) are in good agreement with the experimental data~\cite{CMS:2022dwd,ATLAS:2022vkf}. 
The same is true, however, also for many scenarios of
physics beyond the SM (BSM),
which are motivated by the open questions and shortcomings of the SM.

While no conclusive sign of BSM physics has been discovered so far, extended scalar sectors, featuring parameter regions that are in agreement with all experimental and theoretical constraints, 
are particularly appealing in this context.
Scalar particles play a fundamental role in the proposed answers to several open issues of the SM.
In this regard, the determination of the shape of the Higgs potential is crucial 
for a better understanding of
electroweak symmetry breaking~\cite{Djouadi:1999gv,Djouadi:1999rca}
and of the thermal history of the universe.  
The current knowledge of the Higgs potential,
which in the case of an extended Higgs sector is a complicated function of the components of all involved scalar fields,
is limited to the 
distance in field space of the electroweak vacuum from the origin,
given by the vacuum expectation value (vev), $v\approx 246 \gev$, and the curvature around it, given by the mass of the 
detected Higgs boson of 
about $125 \gev$. The information gathered on the trilinear Higgs coupling (THC) is, however, 
insufficient so far to determine whether a BSM Higgs sector is realized in nature.

Among the most prominent shortcomings of the SM is its inability to explain
the observed baryon asymmetry of the universe (BAU)~\cite{WMAP:2012fli}. 
A dynamical explanation is given by electroweak baryogenesis 
(EWBG)~\cite{Kuzmin:1985mm,Cohen:1990it,Cohen:1993nk,Quiros:1994dr,Rubakov:1996vz,Funakubo:1996dw,Trodden:1998ym,Bernreuther:2002uj,Morrissey:2012db}, 
provided the three Sakharov conditions~\cite{Sakharov:1967dj} are fulfilled. Among these is the departure from thermal
equilibrium. A large barrier in the Higgs potential at the electroweak (EW) phase transition, 
which can arise from
a sizable THC, enables
a strong first order EW phase transition (SFOEWPT), and thus helps to facilitate 
baryogenesis~\cite{Kanemura:2004ch, Noble:2007kk, Huang:2015tdv, Reichert:2017puo}.
Accordingly, the realization of an SFOEWPT is often correlated with a significant enhancement (of at least 20-30\%)
of the THC of the detected Higgs boson, \lahhh, 
compared to the SM prediction~\cite{Kanemura:2004ch,Grojean:2004xa,Basler:2017uxn,Basler:2019iuu}. 
The contributions giving rise to an SFOEWPT and a shift in the prediction for $\lahhh$ can generically occur
in models with extended Higgs sectors via the higher-order corrections involving
additional heavy states~\cite{Kanemura:2004mg, Kanemura:2004ch}. 
It has been demonstrated that in simple extensions such as the Two Higgs Doublet Model (2HDM) 
the loop corrections to $\lahhh$ can change the tree-level 
value by several 100\% while being in agreement with all existing experimental and theoretical constraints~\cite{Kanemura:2004mg,Bahl:2022jnx}. 
Therefore already the present experimental information on 
\lahhh\ (see below) provides an important test of the allowed parameter 
space~\cite{Bahl:2022jnx}. 

Defining by $\kala$ the coupling modifier 
relative to
the tree-level THC in the SM, 
\begin{eqnarray}
\kala \equiv \frac{\lahhh}{\laSM^{(0)}} \;,
\end{eqnarray}
with
\begin{eqnarray}
\laSM^{(0)}= \frac{\Mh^2}{2v^2}\simeq 0.13 \;,
\end{eqnarray}

While the current sensitivity is still far from the SM (tree-level) value of $\kala=1$, 
the existing limits already probe large  
deviations from this value that can occur  
in simple extensions of the SM Higgs sector such as the 2HDM~\cite{Bahl:2022jnx}.

The Large Hadron Collider in its High Luminosity phase (HL-LHC) will be able 
to significantly improve the sensitivity to
possible BSM scenarios~\cite{deBlas:2019rxi}. Current prospects for the sensitivity at the HL-LHC with $3\,{\rm ab}^{-1}$ integrated luminosity per detector are
$-0.5 < \kala < 1.6$ at the $1\sigma$ level in the combination of the $b\bar{b}b\bar{b}$,  
$b\bar{b}\gamma\gamma$ and $b\bar{b}\tau^+\tau^-$ channels~\cite{ATLAS:2022faz}. 
This on the one hand motivates precise theoretical predictions for the Higgs pair production process, which at the (HL-)LHC is dominantly given by gluon fusion into Higgs pairs, taking into account the possibility of sizable BSM contributions to the occurring trilinear Higgs couplings. On the other hand it is important to ensure that the obtained experimental bounds on the gluon fusion Higgs pair production process can be confronted in a meaningful way with theoretical predictions in different scenarios of electroweak symmetry breaking, where a resonant contribution from the exchange of a heavy neutral Higgs boson might be possible in addition to the non-resonant contributions that are always present.  The latter contain in particular a contribution involving the Higgs boson at 125 GeV and a top-loop induced contribution where no Higgs boson enters at leading order. 

In this paper we 
adopt the well motivated 2HDM as theoretical framework, but we stress that our qualitative results are applicable to a wide class of extended Higgs sectors.
We will investigate in particular the effects of two 
contributions entering the process of gluon fusion into Higgs pairs, 
$gg \to hh$, which provides direct access to $\lahhh$ at the LHC.
We will study
the impact of the inclusion of a possible resonant heavy Higgs 
contribution
with subsequent decay into a pair of the Higgs boson at 125 GeV,
involving the trilinear Higgs coupling $\lahhH$ at lowest order. 
Furthermore, we will investigate the effect of potentially large higher-order corrections to $\lahhh$ and $\lahhH$ on the Higgs pair production process. We will demonstrate that the combination of the two effects has important implications on the experimental limits that can be extracted from the Higgs pair production process.

Our paper is organized as follows.~In \refse{sec:2hdm} we describe the 2HDM in more detail and summarize the predictions for the trilinear 
Higgs couplings at tree level (\refse{sec:treethc}) and at one-loop order (\refse{sec:loopthc}). We briefly review 2HDM 
Higgs pair production at the LHC in \refse{sec:hh_2hdm}. We present our results for the 
differential cross section w.r.t.~the invariant mass in \refse{sec:mhh}. 
The confrontation of the Higgs pair production process with 
experimental limits 
is discussed in \refse{sec:experiment}, focusing
on the case of non-resonant production in \refse{sec:nores} and on the case 
of resonant production in \refse{sec:res}. Our conclusions are given in \refse{sec:conclusions}.


\section{Trilinear Higgs Couplings in the 2HDM}
\label{sec:2hdm}

\subsection{Model details}

One of the simplest scalar extensions of the SM is the addition of one complex doublet under the SU(2) symmetry,
resulting in the 2HDM. For simplicity, we assume a $\cp$-conserving
2HDM~\cite{TDLee,Gunion:1989we,Aoki:2009ha,Branco:2011iw}. 
The tree-level scalar potential with a $\mathbb{Z}_2$ symmetry, under which the two complex 
Higgs doublet fields transform as $\Phi_1 \to \Phi_1$ and $\Phi_2 \to -\Phi_2$, is given by
\begin{eqnarray}
V &=& m_{11}^2 (\Phi_1^\dagger\Phi_1) + m_{22}^2 (\Phi_2^\dagger\Phi_2)
- \msq (\Phi_1^\dagger \Phi_2 + \Phi_2^\dagger\Phi_1)
+ \frac{\la_1}{2} (\Phi_1^\dagger \Phi_1)^2 +
\frac{\la_2}{2} (\Phi_2^\dagger \Phi_2)^2 \nonumber \\
&& + \la_3
(\Phi_1^\dagger \Phi_1) (\Phi_2^\dagger \Phi_2) + \la_4
(\Phi_1^\dagger \Phi_2) (\Phi_2^\dagger \Phi_1) + \frac{\la_5}{2}
[(\Phi_1^\dagger \Phi_2)^2 +(\Phi_2^\dagger \Phi_1)^2]  \;,
\label{eq:scalarpot}
\end{eqnarray}
with all coupling and mass parameters being real. 
The $\mathbb{Z}_2$ symmetry is softly broken by the parameter $\msq$.
The fields $\Phi_1$ and $\Phi_2$ can be conveniently parametrized as
\begin{eqnarray}
\begin{split}
\Phi_1 &= \left( \begin{array}{c} \phi_1^+ \\ \frac{1}{\sqrt{2}} (v_1 +
    \rho_1 + i \eta_1) \end{array} \right) \;, \quad \,\,\,\,\,\Phi_2 &= \left( \begin{array}{c} \phi_2^+ \\ \frac{1}{\sqrt{2}} (v_2 +
    \rho_2 + i \eta_2) \end{array} \right) \;,
\label{eq:2hdmvevs}
\end{split}
\end{eqnarray}
in terms of their respective vacuum expectation values, $v_1$ and $v_2$
(with $\sqrt{v_1^2 + v_2^2} \equiv v$), 
and the interaction fields 
$\phi_{1,2}^\pm$, $\rho_{1,2}$ and $\eta_{1,2}$ that mix to give rise to five physical scalar fields and 
three (would-be) Goldstone bosons. The physical fields comprise two $\cp$-even fields,
$h$ and $H$, where by convention $m_h < m_H$, and we identify $h$ with the scalar boson observed at the LHC at about 
$125 \gev$,
one $\cp$-odd field, $A$, and one charged Higgs pair, $H^\pm$. 
The mixing matrices diagonalizing the $\CP$-even and $\CP$-odd/charged Higgs mass matrices can be expressed in 
terms of the mixing angles $\alpha$ and $\beta$, respectively, with $\TB \equiv v_2/v_1$ \footnote{We use the short-hand notation $s_x \equiv \sin x$, $c_x \equiv \cos x, \TB \equiv \tb$.}. 
The ``alignment limit''~\cite{Gunion:2002zf}
corresponds to
$\CBA \to 0$, where the light Higgs boson $h$ has couplings to fermions and gauge bosons at lowest order that exactly correspond to the ones in the SM.

The occurrence of tree-level flavor-changing neutral currents (FCNC) is
avoided by extending the $\mathbb{Z}_2$ symmetry to the Yukawa sector. 
This results in four variants of the 2HDM, 
depending on the $\mathbb{Z}_2$ parities of the 
fermion types. In this article we focus on the Yukawa type~I, where all fermions couple to $\Phi_2$. 
The couplings of the Higgs bosons to SM particles are
determined by the mixing in the
scalar sector.  The couplings of the neutral $\cp$-even Higgs bosons to
fermions are given by

\begin{equation}
	\mathcal{L} =-\sum_{f=u,d,l}\frac{m_f}{v}\left[\xi_h^f\bar{f}fh + \xi_H^f\bar{f}fH \right],
\end{equation}

\noindent 
where $m_f$ are the fermion masses, and $\xi_{h,H}^f$ are the fermionic Yukawa coupling modifiers, which express
the couplings relative to the ones of the SM Higgs. They
are equal for all three generations of up-type quarks ($u$), down-type quarks ($d$) and leptons ($l$). 
In the type~I 2HDM the coupling modifiers are equal for all fermions 
and given by $(f = t,b,\tau)$, 
\begin{eqnarray}
\xi_{h}^{f} = \SBA + \CBA \cot \be, \quad \xi_{H}^{f} = \CBA - \SBA \cot \be.
\end{eqnarray}
The Yukawa hierarchy 
implies that the Higgs boson couples 
predominantly 
to the top quark ($t$) and to a lesser extent
to the bottom quark ($b$). 

We work in the physical basis of the 2HDM, where the Higgs potential parameters are expressed in terms of a set of parameters given mostly by physical quantities as
\begin{equation}
c_{\be-\al},\; t_\beta,\; v,\;
 \Mh,\; \MH,\;  \MA,\; \MHp,\; \msq . 
\label{eq:inputs}
\end{equation}
Here, $\Mh, \MH, \MA, \MHp$ are the masses of the physical scalars.


\subsection{Tree-Level Trilinear Higgs Couplings in the 2HDM}
\label{sec:treethc}

The generic tree-level THCs
$\la_{h h_i h_j}\tree$ involving at least one Higgs boson $h$ with $\Mh \sim 125 \gev$ are defined
such that the Feynman rules are given by 
\begin{equation}
	\begin{gathered}
		\includegraphics{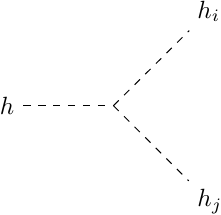}
	\end{gathered}
	= - ivn!\la_{h h_i h_j}\tree~,
\label{eq:lambda}
\end{equation}
where $n$ is the number of identical particles in the vertex.
For our analysis 
in the following the two couplings
$\lahhh$ and $\lahhH$ are relevant.
With the convention given in \refeq{eq:lambda} the 
self-coupling $\lahhh\tree$ has the same normalization at tree-level as
in the SM, where the Feynman rule is given by $-6iv\laSM\tree$. 
The 2HDM tree-level THCs $\lahhh\tree$ and $\lahhH\tree$ can be cast into the forms
\begin{align}
  \lahhh\tree
  &=  -\frac{1}{2} \Big\{ \la_1c_{\be } s_{\al }^3-\la _2 c_{\al }^3 s_{\be }
        +\left(\la_3+\la _4+\la_5\right) \left(c_{\al }^2 c_{\be }
        s_{\al }-c_{\al } s_{\al }^2 s_{\be}  \right) \Big\} \nonumber \\
\label{eq:hhh_phys}
        &= \frac{1}{2v^2} \bigg\{ \Mh^2 s_{\be -\al }^3 
        +\left(3 \Mh^2-2 \bar{m}^2\right) c_{\be -\al }^2 s_{\be -\al }
        +2 \cot 2 \be \left( \Mh^2-\bar{m}^2\right) c_{\be -\al
        }^3\bigg\}, \\[.5em]
\notag
 \lahhH\tree 
 &= \frac{1}{2} \Big\{  3 \la _1  c_{\al } c_{\be } s_{\al }^2
 + 3 \la _2 c_{\al}^2 s_{\al } s_{\be } \\
&\quad\qquad +\left(\la_3+\la _4+\la_5\right)
 \left(c_{\al }^3 c_{\be }-2 c_{\al }^2 s_{\al }
   s_{\be }-2 c_{\al } c_{\be } s_{\al }^2+s_{\al }^3
   s_{\be}\right)\Big\} \nonumber \\
\label{eq:hhH_phys} \notag
        &= -\frac{\CBA}{2v^2} \bigg\{  \left(2\Mh^2+\MH^2-4\mbar\right)\SBA^2
        +2\cot{2\be}\left(2\Mh^2+\MH^2-3\mbar\right) \SBA\CBA \\
        &\quad\qquad -\left(2\Mh^2+\MH^2-2\mbar \right) \CBA^2
         \bigg\},  
\end{align}
where $\bar{m}^2$ is defined as
\begin{eqnarray} 
\bar{m}^2&=&\frac{\msq}{s_\be c_\be} \,.
\label{eq:mbar}
\end{eqnarray} 
From the latter expressions one can easily read off the THCs
in the alignment limit where $\CBA = 0$, namely 
$\lahhh\tree = \laSM\tree$ and
$\lahhH\tree = 0$. 
Away from the alignment limit the predictions for these couplings in the 2HDM, even at tree level, can be significantly modified,
see~e.g.~\citeres{Arco:2020ucn,Abouabid:2021yvw,Arco:2022xum} for 
studies in all four Yukawa types.


\subsection{Loop-Corrected Trilinear Higgs Couplings}
\label{sec:loopthc}

In the 2HDM, it has been shown that 
the loop contributions to the THCs involving the heavy BSM Higgs bosons can give rise to
corrections of the order of 100\% and larger~\cite{Kanemura:2002vm,Kanemura:2004mg} w.r.t.\ their tree-level
values. More recently, also two-loop corrections have been computed~\cite{Braathen:2019pxr} enhancing in 
some parts of the parameter space the value of $\kala$ to the sensitivity of current and future runs of the
LHC~\cite{Bahl:2022jnx}. The occurrence of large loop corrections should, however, not been regarded  
as a sign of the breakdown of perturbation theory, as large corrections 
at one-loop order are present mainly due to new contributions involving couplings of the Higgs boson $h$ to heavier BSM Higgs bosons 
that do not appear at tree level~\cite{Kanemura:2004mg}, while
the size of the two-loop corrections relative to 
the one-loop result follows the expected perturbative 
behavior~\cite{Braathen:2019pxr, Braathen:2019zoh, Bahl:2022jnx}. In view of these findings 
the impact of these large higher-order corrections
on the Higgs pair production process should be studied.

For the computation of the one-loop corrections to the THCs contributing to
our numerical analysis
we use the public code \texttt{BSMPT}~\cite{Basler:2018cwe,Basler:2020nrq}, where the trilinear Higgs couplings are 
extracted from the one-loop corrected effective potential (evaluated here at zero temperature), 
\begin{equation}
    V_{\rm eff} = V_{\rm tree} + V_{\rm CW} + V_{\rm CT}~.
\end{equation}
In this equation, $V_{\rm tree}$ is the tree-level potential of the 2HDM given in \refeq{eq:scalarpot}, $V_{\rm CW}$ 
is the one-loop Coleman--Weinberg potential~\cite{Weinberg:1973am,Coleman:1973jx} at zero temperature, and $V_{\rm CT}$ is the 
counterterm potential. The counterterm potential is chosen such that the masses and mixing angles are kept at 
their tree-level values, which therefore allows us to conveniently use them as inputs in our scans.
In this set-up, the ``effective loop-corrected trilinear Higgs couplings" can be computed as the third derivatives of 
the effective potential with respect to the Higgs fields, evaluated at the minimum,
\begin{eqnarray}
    \lahhh\one = \frac{1}{3!v} \frac{\partial^3 V_{\rm eff}}{\partial h^3}\bigg|_{h,H=0},\,\,\,\,\, \lahhH\one = \frac{1}{2!v} \frac{\partial^3 V_{\rm eff}}{\partial h^2 \partial H}\bigg|_{h,H=0}. 
\end{eqnarray}
Alternatively, one could use a fully diagrammatic approach by calculating the one-loop corrections with the public tool \texttt{anyH3} \cite{Bahl:2023eau}, where the extension of the provided results for $\lahhh\one$ to $\lahhH\one$ and further trilinear Higgs couplings is currently under development.


\section{Higgs Pair Production in the 2HDM}

\subsection{Theoretical introduction}
\label{sec:hh_2hdm}

\begin{figure}[ht!]
  \begin{center}	
  \includegraphics[width=1\textwidth]{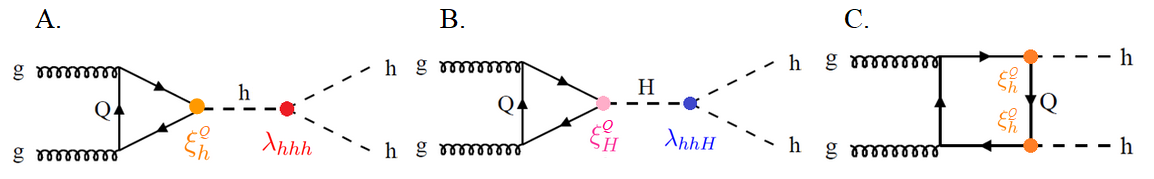}
  \end{center}
\caption{Generic diagrams contributing to the
pair production 
of the Higgs boson $h$ at about 125~GeV
in gluon fusion within the 2HDM, mediated by 
heavy quark loops, $Q = t,b$. The red and blue dots denote the triple Higgs couplings $\lahhh$ and $\lahhH$,
respectively, evaluated at leading or next-to-leading order; the orange (pink) dot denotes the $h$ ($H$) 
Yukawa coupling parametrized by the coupling modifier $\xi_h^Q$ ($\xi_H^Q$). The diagrams labelled as A and C are the
continuum diagrams, which appear in analogous form in the SM. The diagram labelled
B is the resonant diagram, involving the $s$-channel heavy $H$ exchange.}
\label{fig:gghhdiagrams}
\end{figure}

The trilinear Higgs boson self-coupling is directly accessible in Higgs pair production. At the LHC, the dominant process is
gluon fusion into Higgs pairs, which at leading order is mediated by heavy quark loops, 
see~\reffi{fig:gghhdiagrams}. The bottom quark contribution in the SM only plays a subleading role, 
whereas in the 2HDM it can be enhanced by large values of $\TB$, depending on the Yukawa type.
The THCs enter through the $s$-channel diagrams, as shown in the first two diagrams of \reffi{fig:gghhdiagrams}.
In the SM, the triangle and box diagrams interfere destructively leading to a relatively small cross section of 
$\sigma_{\rm SM} = 31.05^{+6\%}_{-23\%}$~fb  at 13 TeV center-of-mass energy \cite{Grazzini:2018bsd,Baglio:2020wgt}\footnote{This is the value 
obtained at NNLO\_FTapprox 
for $m_h=125$~GeV with the renormalization and factorization scale chosen to be half the invariant Higgs pair mass for a c.m.~energy of $\sqrt{s} = 13$ TeV~\cite{Grazzini:2018bsd}. 
At NNLO\_FTapprox, the cross section is computed at next-to-next-to-leading order (NNLO) QCD in the heavy-top limit \cite{deFlorian:2013jea,Grigo:2014jma,Davies:2021kex} with full leading order (LO) and next-to-leading order (NLO) mass effects \cite{Borowka:2016ehy,Borowka:2016ypz,Baglio:2018lrj,Baglio:2020ini,Baglio:2020wgt} and full mass dependence in the one-loop double real corrections at NNLO QCD. The uncertainty combines the uncertainty from the renormalization and factorization scale variations with the uncertainty due to the choice of the renormalization scheme and scale of the mass of the top quark~\cite{Baglio:2020wgt}.}. The LO QCD value at 13 TeV, which is the energy we use for all the plots shown in the paper, is $\sigma_{\rm SM} = 16.6$~fb, and the NLO QCD value in the Born improved heavy-top limit is 32.7 fb.

In the 2HDM, there are two potential sources of changes w.r.t.\ the SM. 
Firstly, the couplings in the SM-like diagrams can differ from the SM values. Whereas the top-Yukawa coupling is restricted by the current constraints to about $\pm$10\% the SM value, much larger deviations
in the trilinear Higgs self-coupling $\lambda_{hhh}$ 
are possible in accordance
with all relevant constraints \cite{Abouabid:2021yvw,Bahl:2022jnx}. Changes in $\lambda_{hhh}$ can 
modify the interference of the SM-like triangle and box diagrams.
Secondly, there is an additional $s$-channel contribution from the heavy Higgs boson, involving the trilinear coupling 
\lahhH\ and the top Yukwawa coupling of the $H$. In case the mass $m_H$ exceeds twice the  mass of the lighter Higgs boson,
$m_H > 2 m_h \sim 250 \gev$, this contribution can lead to resonant $hh$ production, in which case the corresponding diagram is referred
to as ``resonant diagram''. Thereby, the cross section can be significantly enhanced. 
On the other hand, depending on the involved couplings and masses, there can also be destructive interferences between the triangle diagrams of the $h$ and $H$ exchange and the box diagram.
Accordingly, the loop contributions to the trilinear Higgs couplings are expected to have an important impact both on the prediction for the
inclusive cross section and
also for the shape of the invariant mass distributions, as will be discussed in the next section.

In this work, we include for the first time in the 2HDM\footnote{For investigations of the effect in the SM, see \citere{Muhlleitner:2022ijf}, and in the next-to-minimal supersymmetric extension of the SM (NMSSM), see \citeres{Nhung:2013lpa,Borschensky:2022pfc}.} the one-loop corrections to the triple Higgs
couplings in the computation of Higgs pair production and analyze their effects. 
It should be noted that in the effective trilinear coupling approach, as defined above, the couplings are evaluated in the approximation of vanishing external momenta. Taking into account the appropriate momentum dependence for the Higgs pair production process would be expected to modify the predictions for
the total di-Higgs production cross section 
only at the percent level in the 2HDM type~I~\cite{Bahl:2023eau}.
Furthermore, the loop-corrected effective trilinear couplings 
constitute the leading contributions to
the full EW corrections for scenarios in which the loop corrections to \lahhh\ and/or \lahhH\ are very large. In this case, contributions beyond the trilinear Higgs self-couplings,
e.g.\ including additional powers of the top Yukawa couplings,
can be shown to be sub-dominant~\cite{Bahl:2022jnx}.
Therefore, for the case of sizable loop corrections to the THCs our results should provide a good approximation to the full electroweak loop corrections to the inclusive process at this order. 

In regions where these corrections are relatively small, which for the non-resonant case implies that the predicted cross sections are significantly below the current experimental sensitivity,
this approach becomes less accurate and a complete next-to-leading-order (NLO) electroweak (EW) calculation of the cross section
would be required, which is 
beyond the scope of this work.\footnote{
For results on the NLO EW corrections to SM Higgs pair production, see \citeres{Muhlleitner:2022ijf,Davies:2022ram,Davies:2023npk,Bi:2023bnq}.
} The aim of our work, on the other hand, is an analysis of
possible implications of large loop contributions and interference effects, in particular regarding the interpretation of the experimental results.
For this purpose the approximate approach pursued here 
should be sufficiently accurate. 

For the numerical evaluation, we use
the code {\tt HPAIR}~\cite{Plehn:1996wb,Dawson:1998py,Grober:2017gut,Abouabid:2021yvw}, adapted to the 2HDM.
This code was originally designed to compute within the SM and its Minimal Supersymmetric Extension (MSSM) the cross
sections for the production of two neutral Higgs bosons through gluon fusion at the LHC. 
The calculations are carried out at leading order (LO) with the full top-quark mass dependence
and include NLO QCD corrections, assuming the limit of an infinite top-quark mass and neglecting bottom loop
contributions.\footnote{Recently, the full top-quark mass dependence at NLO QCD has been provided for the production of an $hH$ pair 
as well as for a $\CP$-odd Higgs pair in the 2HDM \cite{Baglio:2023euv}.} However, in the 2HDM the latter assumption can become less accurate at large values of $\TB$
due to the increased importance of the bottom quark loop contribution, depending on the Yukawa type. In the 
Yukawa type~I, which is used throughout our analysis, no enhancement of the bottom Yukawa coupling occurs. 
Furthermore, for this analysis, we have modified {\tt HPAIR} to include the one-loop corrections to the THCs 
as described in \refse{sec:loopthc}. The PDF set used in the numerical evaluation at LO (NLO) QCD is CT14lo (CT14nlo) \cite{Dulat:2015mca}.


\subsection{Impact of loop corrections to the trilinear Higgs couplings on invariant mass distributions}
\label{sec:mhh}

\begin{figure}[ht!]
\begin{center}
\includegraphics[width=0.8\textwidth]{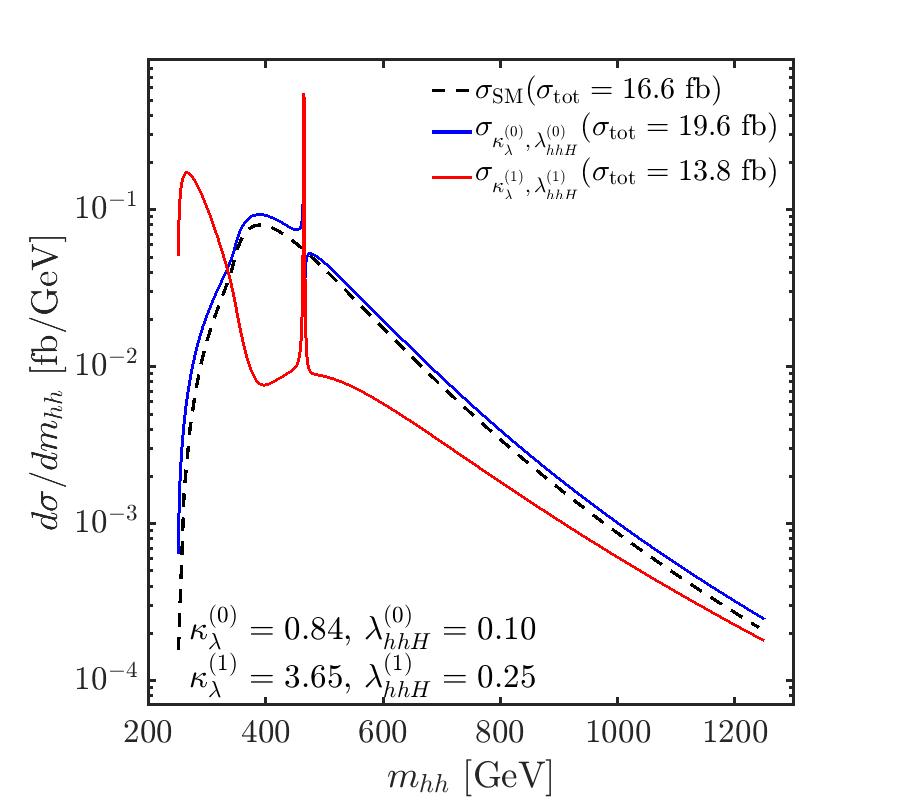}
	\end{center}
\caption{Invariant mass distribution for the benchmark point in the 2HDM type~I defined in \refeq{eq:bp_lonlo}. 
The SM prediction
(dashed
black line) 
is shown together with the 2HDM results with (solid red line) and without (solid blue line) 
loop corrections to the THCs,
see text.
}
\label{fig:thcgreen}
\end{figure}

In this section, we explore the behavior of the invariant mass distribution of the di-Higgs final state 
when incorporating loop corrections to the THCs involved in Higgs pair production. 
In \reffi{fig:thcgreen} we present various \mhh\ distributions for a sample benchmark point in the 2HDM of type~I. It is defined by the input parameters

\begin{eqnarray}
\begin{split}
\TB &= 10,\; \CBA = 0.13\; (\SBA>0)\; \\ \MH = 465\gev, \; \MA &= \MHp = 660 \gev \; \mbox{ and } \msq = \MH^2\CA^2/\TB.
\label{eq:bp_lonlo}
\end{split}
\end{eqnarray}
For this point we find 
\begin{eqnarray}
\kala\tree \equiv \frac{\lahhh\tree}{\laSM\tree}= 0.84, \; \kala\one \equiv \frac{\lahhh\one}{\laSM\tree}= 3.65, \; \lahhH^{(0)} = 0.10 \mbox{ and } \lahhH\one = 0.25,
\label{eq:bpex}
\end{eqnarray}
\KR{
and the rest of the trilinear couplings at tree level are:
\begin{eqnarray}
\lambda_{hHH}^{(0)} = -2.31, \; \lambda_{HHH}^{(0)} = 0.07, \; \lambda_{hAA/hH^{\pm}H^{\pm}}^{(0)} = -5.83 \mbox{ and } \lambda_{HAA/HH^{\pm}H^{\pm}}^{(0)} = 0.54.
\label{eq:thcbpex}
\end{eqnarray}
}

The THC of the SM-like Higgs boson is hence very SM-like at tree level, but substantially increased by one-loop corrections. The THC between the heavy Higgs boson and the two light Higgs bosons is increased by 150\% by the one-loop corrections.

Concerning the invariant mass distributions shown in our analysis, it is important to note that they 
are calculated at leading order. It would be possible to compute the invariant mass spectrum with {\tt HPAIR} at NLO QCD in the Born improved heavy-top limit. However, it has been shown that mass effects may significantly distort the NLO  distributions \cite{Borowka:2016ehy,Borowka:2016ypz,Baglio:2018lrj,Baglio:2020ini,Baglio:2020wgt}. While, for the 2HDM, the full mass effects at NLO QCD have been considered in \citere{Baglio:2023euv}, there exists
no public code that allows us to obtain results for our benchmark scenarios, in particular including resonances. In \citere{Buchalla:2018yce} a parametrisation has been given for the total cross section and the $m_{hh}$ distribution in the framework of a non-linear effective field theory as a function of the anomalous Higgs couplings that includes NLO corrections. While this framework considers deviations from the SM Higgs sector, it however does not include the possibility of additional Higgs bosons. 
Consequently, one has
the choice between a LO distribution ignoring NLO effects and an approximate NLO distribution ignoring finite top-mass 
effects at NLO, where we chose to adopt the LO case. While this approach obviously cannot capture the full NLO mass effects, it 
does provide information regarding the possible impact of a BSM Higgs boson resonance and of 
NLO electroweak corrections to THCs on the \mhh\ distribution, 
which is the main goal of our analysis. 
The inclusive cross section, on the other hand, is rather well approximated at NLO QCD by applying a $K$-factor, obtained from the ratio of the NLO to the LO cross section, of $K(\mathrm{NLO}) \approx 2$~\cite{Arco:2022lai}.

The blue curve in \reffi{fig:thcgreen} is the invariant mass distribution for the specified benchmark point with both THCs taken at tree-level, whereas the red line displays the result for
the distribution for the case where both THCs are incorporated at the one-loop level. The dashed black line indicates the SM prediction. 
Starting our discussion with
the tree-level distribution (blue
line), several features can be noticed. The small values of the differential cross section just above the 
threshold are a consequence
of a cancellation of the form factors involved in the continuum diagrams (diagrams A and C in Fig.~\ref{fig:gghhdiagrams}). 
The invariant mass distribution reaches a maximum
at $\mhh \approx 400 \gev$,
which is related to the di-top on-shell production and is also present in the distribution of single Higgs 
production (see e.g.~\citere{Cepeda:2019klc}). 
A further striking feature is the
resonance located at  $\mhh \approx \MH$ 
showing a peak-dip structure. 
Apart from the resonant contribution, the shape of the tree-level distribution 
resembles the SM prediction (dashed black line), taking into account the \KR{relatively small} value of $\lahhh^{(0)}$.

Turning to the red line, incorporating one-loop corrections to both THCs, one can
observe that the 
shape of the distribution
changes drastically. In particular 
the cancellation close to the kinematical threshold in the leading order distribution is lifted.\footnote{This effect has already been observed in the context of the SM in \citere{Muhlleitner:2022ijf}.} This cancellation now happens at values of
$\mhh \approx 400 \gev$ and 
leads to a large reduction of the differential cross section in the region where
at leading order a maximum occurred. Furthermore, close to the
kinematical threshold the distribution is largely enhanced, leading to the appearance of a 
structure resembling a peak at $m_{hh} \approx 250 \gev$.
We also investigated the impact of the one-loop corrections to the two THCs individually (not shown in the plot) and found 
that in this scenario the corrections to $\lahhH$ play a minor role, while the biggest
changes are 
caused by the large one-loop corrections to $\kala$. 

Also shown in the figure are the total cross section values\footnote{The total cross section values are given at LO QCD in accordance with the distributions given at LO. As stated above, including the NLO QCD corrections obtained with {\tt HPAIR}, the cross section values would increase by about a factor 
of~2~\cite{Arco:2022lai}.}. Here it is interesting to note that the decrease in the tree level value of $\kala$ of about 15\% w.r.t.\ the SM\footnote{Here we use the LO SM total cross section, $\sigma_{\rm SM}$, as given in \refse{sec:hh_2hdm} for comparison.} leads to an increase of roughly 20\%
of the tree level cross section, whereas the inclusion of the one-loop corrections to the THCs results in a 
reduction of the 2HDM cross section by about 30\%, i.e.\ 20\% smaller than the SM result.

\begin{figure}[ht!]
  \begin{center}
\includegraphics[width=0.85\textwidth]{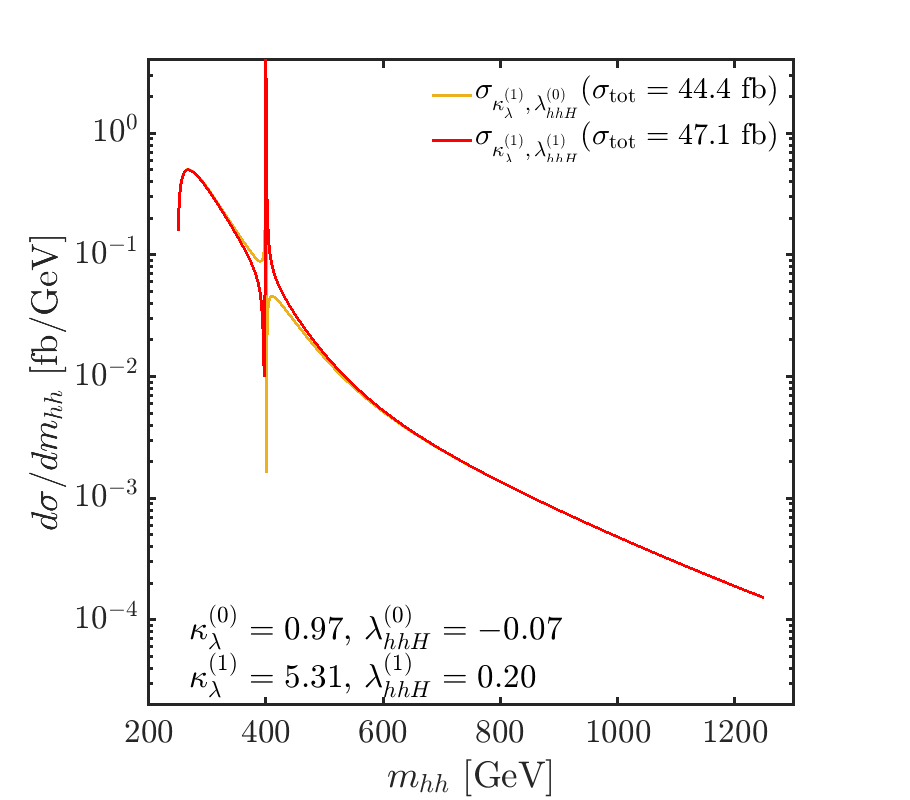}
	\end{center}
\caption{
Impact of the loop corrections to \lahhH\ on the resonance shape for the benchmark point defined in \refeq{eq:bp_112}.
The red (orange) line shows the result with (without) loop corrections to $\lahhH$.}
\label{fig:lahhHmhh}
\end{figure}

In \reffi{fig:lahhHmhh} we present an example where the loop corrections to \lahhH\ play a crucial role. 
The input parameters are 

\begin{eqnarray}
\begin{split}
 t_\beta &= 15, \; \CBA = 0.12\; (\SBA>0),\\ \; \MH = 400 \gev, \; \MA &= \MHp = 660 \gev \mbox{ and } \msq = \MH^2 s_{\beta} c_{\beta} \;.   
 \label{eq:bp_112}
\end{split}
\end{eqnarray}
For these parameters we find 
\begin{eqnarray}
\kala\tree = 0.97, \; \kappa_\lambda^{(1)} = 5.31, \; \lahhH\tree = -0.07 \mbox{ and }\lahhH^{(1)}= 0.20 \;,
\end{eqnarray} 
\KR{
and
\begin{eqnarray}
\lambda_{hHH}^{(0)} = -2.27, \; \lambda_{HHH}^{(0)} = 0.12, \; \lambda_{hAA/hH^{\pm}H^{\mp}}^{(0)} = -6.77 \mbox{ and } \lambda_{HAA/HH^{\pm}H^{\mp}}^{(0)} = 0.67.
\label{eq:thcbpex}
\end{eqnarray}
}
The result including NLO corrections only to $\kala$ is shown as orange solid line, and corresponds to a total LO QCD cross section of $44.4 \fb$. The \mhh\ distribution shows a 
pronounced peak--dip structure
at $\mhh \sim \MH$. The result including the one-loop corrections to both THCs is shown as solid red line.
The incorporation of
the higher-order corrections 
results in a larger $\lahhH^{(1)}$ value with opposite sign
compared to the tree-level value. Its inclusion gives rise to
a dip--peak structure, 
i.e.\ the opposite behavior compared to the
tree-level case. This effect 
is caused by
a change in the overall sign of the couplings involved in the resonant diagram,
$\lahhH \times \xi^t_H$, as discussed in \citere{Arco:2022lai}. In the present example we demonstrate 
that such a change can arise solely from one-loop corrections to \lahhH, i.e.\ the incorporation of electroweak loop 
corrections is crucial in this case for a reliable prediction of the
experimental signature (experimental effects like smearing due to a limited detector resolution will be discussed in the next section). 
This effect is clearly visible even in the case of large 
one-loop corrections to \kala, 
as it is the case
in this example. 
Our discussion
highlights the relevance of higher-order corrections also in the THCs involving BSM Higgs bosons, as they can 
have a drastic effect on the invariant mass distributions.


\section{Confrontation with
experimental limits}
\label{sec:experiment}

In view of the significant improvements in the experimental sensitivity to the  
di-Higgs production cross section 
that have occurred recently and are expected to be achieved in the future
it is crucial that the experimental limits (and of course eventually also the experimental measurements) are presented in such a way that they can be confronted with theoretical predictions in different scenarios of electroweak symmetry breaking in a well-defined way. Up to now the experimental limits presented by ATLAS and CMS are given either for non-resonant production, taking into account only SM-like contributions, or for purely resonant production, where SM-like non-resonant contributions are omitted. We discuss both types of limits in the following. 


\subsection{Non-resonant production}
\label{sec:nores}

\begin{figure}[ht!]
\begin{center}
\includegraphics[width=0.49\textwidth]{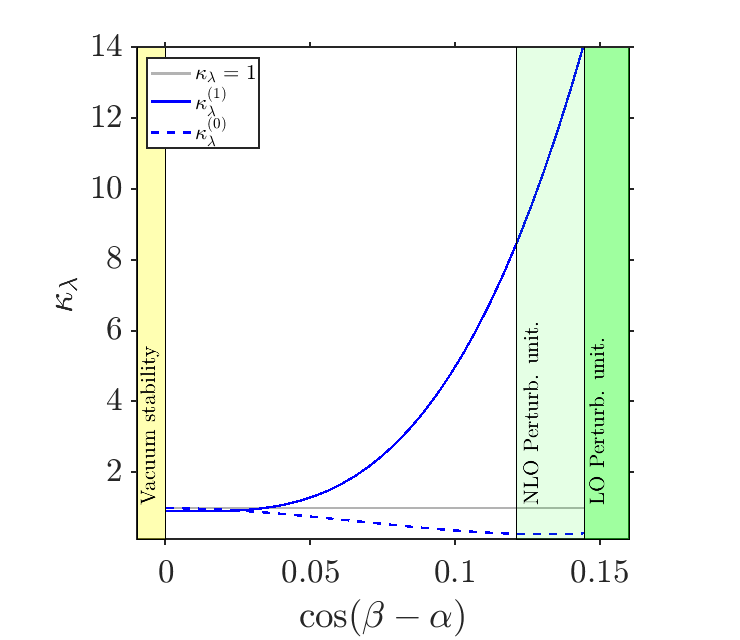}
\includegraphics[width=0.49\textwidth]{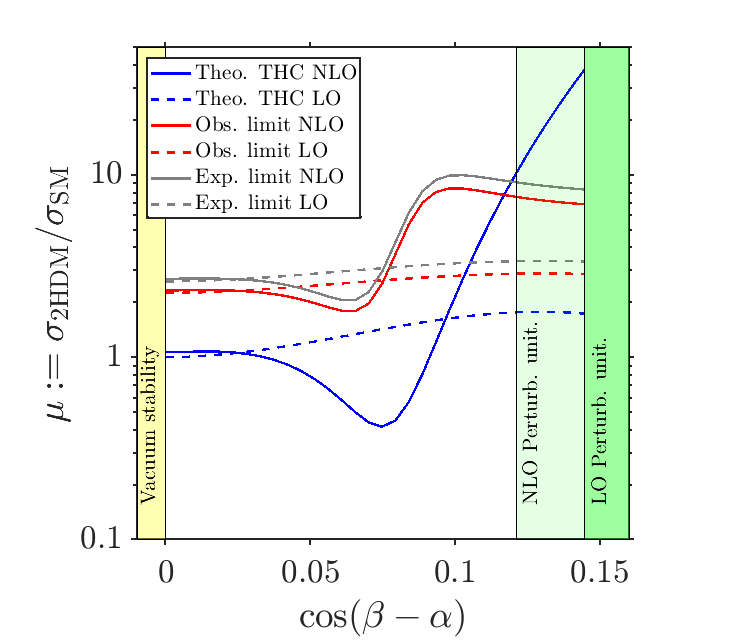}
	\end{center}
\caption{2HDM type I scenario described in \refeq{eq:bps_nores}.
Left: \kala\ as a function of $\CBA$. The gray (blue dashed, blue solid) line shows the result for
$\laSM\tree$ ($\lahhh\tree$, $\lahhh\one$), 
normalized to $\laSM\tree$.
Right: Limits on $\mu \equiv  \sig_\THDM/\sig_\SM$ (each cross section calculated at LO QCD) as function of $\CBA$.
Red, gray and blue: expected, observed experimental limits and theory predictions with \kala\ taken at LO  (dashed) and NLO (full). In both plots the parameter space excluded by theoretical constraints 
is indicated by the yellow (vacuum stability), dark green (perturbative unitarity at LO) and light green 
(perturbative unitarity at NLO) shaded areas.
} 
\label{fig:mu_cba}
\end{figure}

We start our discussion with the analysis of the non-resonant limits. 
In this case the experimental limits
are obtained under the assumption that there is no
contribution from an $s$-channel exchange of an additional Higgs boson, i.e.\ only the contributions of diagrams A and C in \reffi{fig:gghhdiagrams} are taken into account.
The most recent results from  ATLAS~\cite{ATLAS:2022jtk} and 
CMS~\cite{CMS:2022dwd} report a limit on the cross section of $gg \to hh$, 
which depends on the value of $\kala$,
and a bound on $\kala$ is extracted.
This is done by comparing the experimental limit with the SM prediction for a varied $\kala$.
We show in \reffi{fig:mu_cba} an
example of the application of these limits for one particular benchmark
scenario in the 2HDM, where we vary $\CBA$. The chosen input parameters are 
\begin{eqnarray}
\begin{split}
t_\beta &= 10, \; \CBA \in \{ 0 \;\ldots\; 0.16\} \;(\SBA>0), \\ \MH = \MA &= \MHp = 1000 \gev, \;
\msq = \MH^2\CA^2/\TB \;.
\label{eq:bps_nores}
\end{split}
\end{eqnarray}
The large $\MH$ value ensures that the resonant contribution from the $s$-channel $H$ exchange is
negligible
(we do not discuss effects of varying \lahhH\ in this context).
The variation of $\CBA$ results in a variation of \kala\ as indicated in the left plot of \reffi{fig:mu_cba}. 
The blue dashed line shows the prediction for $\kala$ at lowest order, 
while the blue solid line shows the one-loop prediction for $\kala$. The gray line indicates the value of $\kala = 1$, which corresponds to a coupling value 
of $\lahhh = \laSM\tree$. 
The parameter spaces that are excluded by theoretical constraints 
are indicated by the yellow (vacuum stability), dark green (perturbative unitarity at LO) and light green 
(perturbative unitarity at NLO) shaded areas. For the 
application
of these limits we used the public package 
\texttt{thdmTools}~\cite{Biekotter:2023eil}.
The constraints from vacuum stability exclude the displayed yellow region 
with negative values of $\CBA$.
For the largest positive values of $\CBA$ the tightest bound arises from perturbative unitarity
(for the constraints at LO and NLO we require that the eigenvalues of the 
$2 \to 2$ scattering matrix satisfy $|a_0|<1$, where $a_0$ denotes the $s$-wave amplitude of the 
scattering process). 
Demanding that the measured properties
of the Higgs boson at $125 \gev$ 
should be satisfied poses
a bound that is weaker than the one from NLO perturbative unitarity and therefore 
this bound is not explicitly shown in the plot.
It can be observed that at tree~level the variation of $\CBA$ towards larger values 
results in a decrease of $\kappa_\lambda^{(0)}$, which reaches values close to zero for $\CBA \gsim 0.1$. Including the one-loop corrections,
as shown by the blue solid line, yields a strong increase of $\kala\one$, with 
$\kala\one \gsim 5$ for
$\CBA \gsim 0.1$ in this example.

In the right plot we present the corresponding experimental limits and theoretical predictions for the ratio between the 2HDM and SM di-Higgs cross sections,
$\mu \equiv \sig_\THDM/\sig_\SM$, both calculated at LO QCD. The solid (dashed) blue line shows the theory prediction using the one-loop (tree-level)
value for \kala. The dark
red line shows the latest experimental observed limit from non-resonant searches reported by ATLAS \cite{ATLAS:2022jtk}. 
The solid (dashed) line indicates the observed limit 
for the value of $\kala$ that we have calculated
at NLO (LO). 
The corresponding gray line represents the expected limit for \kala\ at NLO (LO).
Confronting the experimental limits with the theoretical predictions,
a value of $\CBA$ is regarded as excluded if the predicted cross section is larger than the experimentally excluded one.
One can see that non-resonant di-Higgs searches would not exclude any value of $\CBA$ for the case where $\kala\tree$ is used. 
As a consequence of the large loop corrections to $\kala$
this changes once the one-loop corrections are taken into account. 
One can see that in this case for the considered example
the non-resonant searches
exclude a region 
for large $\CBA$ values that is 
allowed by all 
other constraints. 
This underlines the fact that the
search for di-Higgs production at the LHC 
already provides sensitivity to parameter regions of the 2HDM that were 
unconstrained so far, see also \citere{Bahl:2022jnx}, where scenarios with 
$\CBA = 0$ have been considered.


\subsection{Resonant production} 
\label{sec:res}

We now turn to the interpretation of experimental limits for resonant di-Higgs production in the 2HDM.
The resonant limits that have been presented by ATLAS and CMS so far were obtained assuming that only one heavy resonance is realized, neglecting the contributions of the continuum diagrams. 
This approach is potentially problematic since in any realistic scenario
the contributions of the non-resonant diagrams A and C in 
\reffi{fig:gghhdiagrams} will of course always be present in addition to the 
possible resonant contribution of an additional Higgs boson. The limits obtained by ATLAS and CMS can therefore only be directly applied to scenarios where the impact of the non-resonant diagrams A and C in 
\reffi{fig:gghhdiagrams} is negligible compared to the contribution of the resonant diagram B. Using the 2HDM as a test case for scenarios that have been claimed to be excluded or non-excluded by ATLAS and CMS we will investigate in the following to what extent the assumption made in obtaining the experimental limits is justified. 

We note that the assumption 
of restricting to the resonant contribution implies that the \mhh\ distribution corresponding to the assumed signal will have 
a peak structure located at $\mhh \approx \MH$.
This peak structure can potentially be modified by the continuum contributions and by interference effects, where the latter in particular can give rise to peak--dip or dip--peak structures.
In the context of assessing the non-resonant contribution arising from the exchange of the detected Higgs boson at 125~GeV (diagram A in \reffi{fig:gghhdiagrams}) we will analyze the impact of loop corrections to $\kala$.   

\begin{figure}
  \begin{center}
\includegraphics[width=0.49\textwidth]{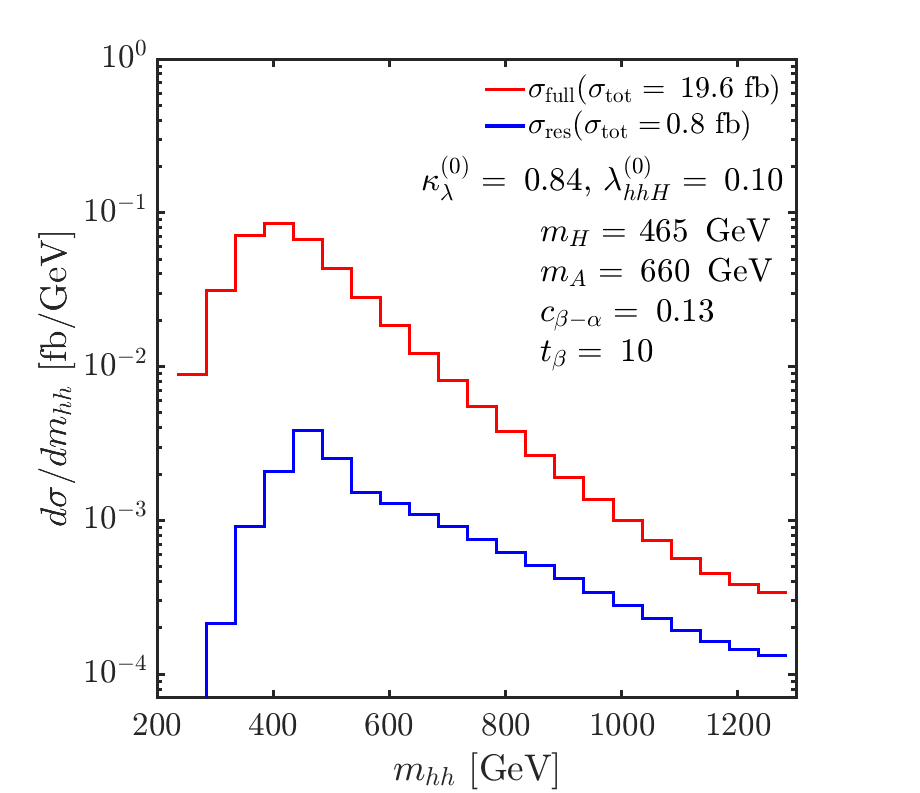}
\includegraphics[width=0.48\textwidth]{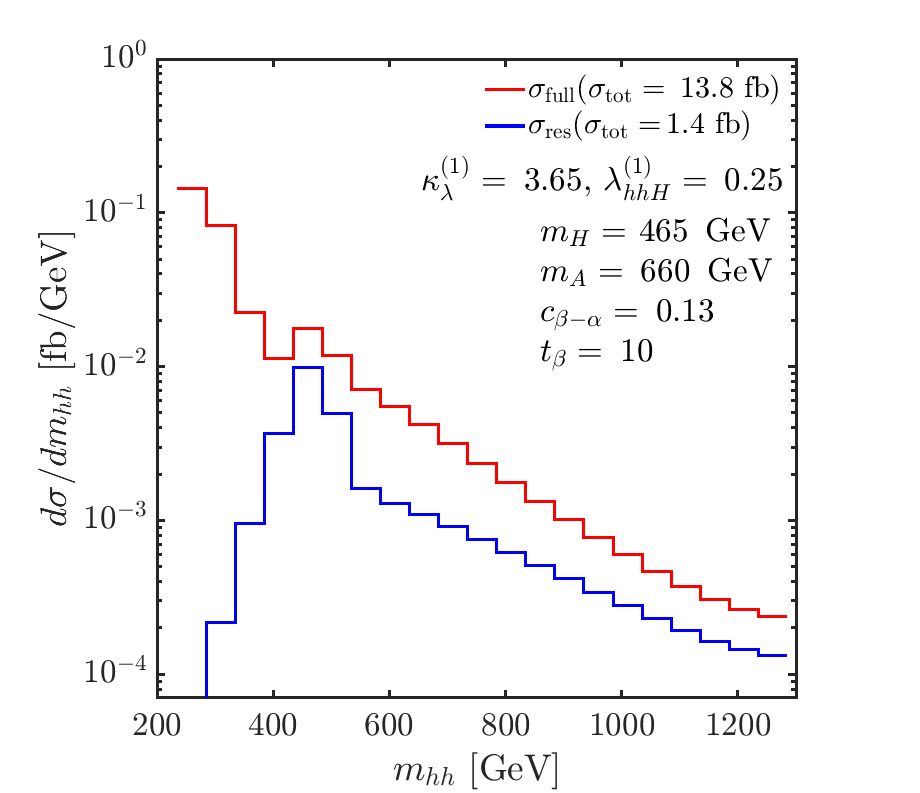}
	\end{center}
\caption{Invariant mass distribution for the 2HDM type I benchmark point defined in \refeq{eq:bp_lonlo}.
Left (right) plot: using $\kala\tree$, $\lahhH\tree$ ($\kala\one$, $\lahhH\one$). Red (blue):  Complete $\sigma(gg \to hh)$ prediction (resonance contribution only).} 
\label{fig:mhhlimits}
\end{figure}

As a first step, to demonstrate the various possible interference and higher-order effects,
we show in \reffi{fig:mhhlimits} the invariant mass distributions for the benchmark point used in \reffi{fig:thcgreen}, which is defined in \refeq{eq:bp_lonlo}. This benchmark point is allowed by all theoretical and experimental constraints.
The blue curves show the pure resonant result, while the red curves correspond to the complete model calculation, including also the non-resonant diagrams and the interference contributions.
The left (right) plot uses the THCs at LO (NLO). Their values and those of the corresponding total cross sections are specified in the plots. 
Contrary to the plots in the previous subsections, here we apply a smearing 
of 15\% and a binning in \mhh\ of $50 \gev$
in order to take into account the limited detector resolution in the experimental analyses, see \citere{Arco:2022lai} for details.

For the case where the tree-level THCs are used, as shown in the left plot of \reffi{fig:mhhlimits}, one can observe that 
the peak in the \mhh\ distribution 
given by the pure resonant distribution 
is 
broadened substantially over several \mhh\ bins
as a consequence of
the inclusion of the non-resonant 
contributions. The effect of the resonance itself is very small, since its contribution to the full result is only about 4\%. Furthermore, the ``resonance-like'' structure of the full result is caused dominantly by the contribution of the continuum diagrams, which peaks slightly above the di-top production threshold ($\sim 400 \gev$), while the resonant contribution (at $\sim 465 \gev$) in this case is minor and does not appear as a clear resonant structure above the continuum distribution.
As can be inferred from the right plot, 
the inclusion of the NLO contributions to the THCs enhances the pure resonant distribution 
in this example
due to the increased absolute size of $\lahhH\one$ in 
comparison with $\lahhH\tree$, which is also reflected in the result for the total cross section.
As indicated by the red curve in the right plot, the combined effect of taking into account non-resonant contributions, interference effects and the NLO corrections to the THCs has a drastic effect on the predicted \mhh\ distribution.
Instead of a pronounced peak as it would be expected from the pure resonant contribution, the full result incorporating all relevant contributions gives rise to an \mhh\ distribution that is overall smoothly falling with just a small modulation near $\mhh \approx \MH$. 
Resolving this structure experimentally will clearly be much more challenging than it would be the case if the distribution had the form obtained from restricting to the pure resonant contribution.
A striking feature that can be inferred from the plot is the large effect of the non-resonant contributions on
the \mhh\ distribution just above the threshold
at $\mhh \sim 250 \gev$. In this region the differential cross section for the full result differs by several orders of magnitude from the one for the pure resonant contribution.
The shape of the differential cross section in this region is also very significantly
modified in comparison to the prediction using the THCs at lowest order (red curve in the left plot). As discussed above, the latter
large enhancement
happens as a result of a the change in \kala\, which 
affects the cancellation between the triangle and box form factors of the continuum diagrams that is present at the \mhh\ threshold at leading order. For \kala\ $\neq$ 1 this cancellation does not take place, giving rise to a large 
enhancement just above the threshold.

\begin{figure}[ht!]
  \begin{center}
\includegraphics[width=0.85\textwidth]{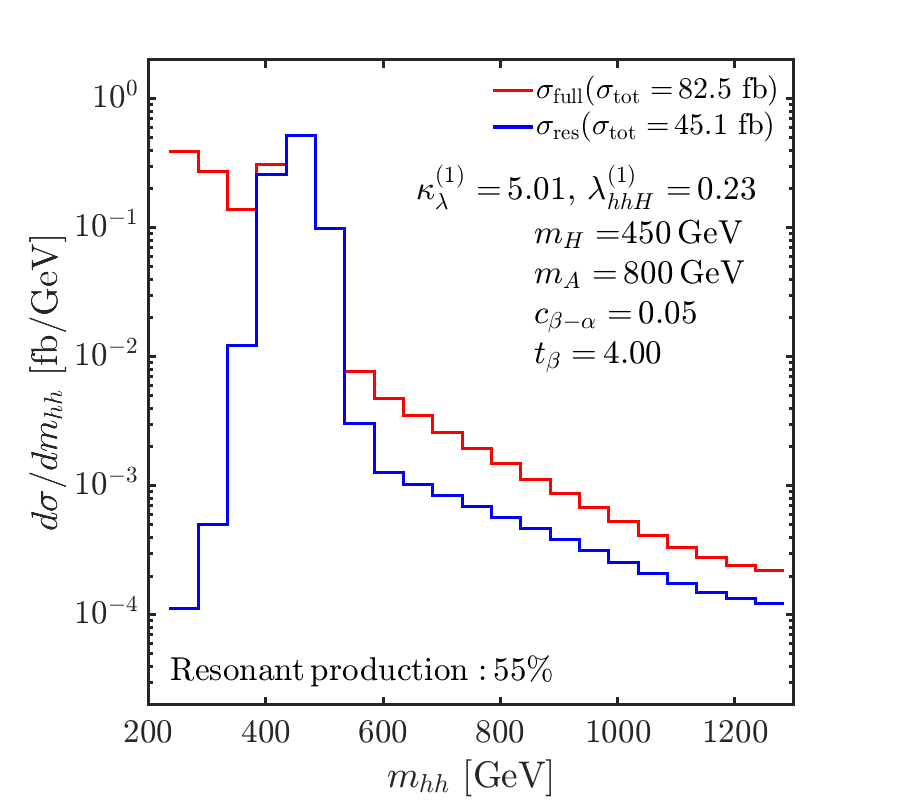}
	\end{center}
\caption{BP1 (allowed by non-resonant searches, excluded by resonant searches): Invariant mass distribution versus the invariant mass for the full result (red) and the result based on the pure resonant contribution (blue).} 
\label{fig:bp1}
\end{figure}

\begin{figure}[ht!]
  \begin{center}
\includegraphics[width=0.85\textwidth]{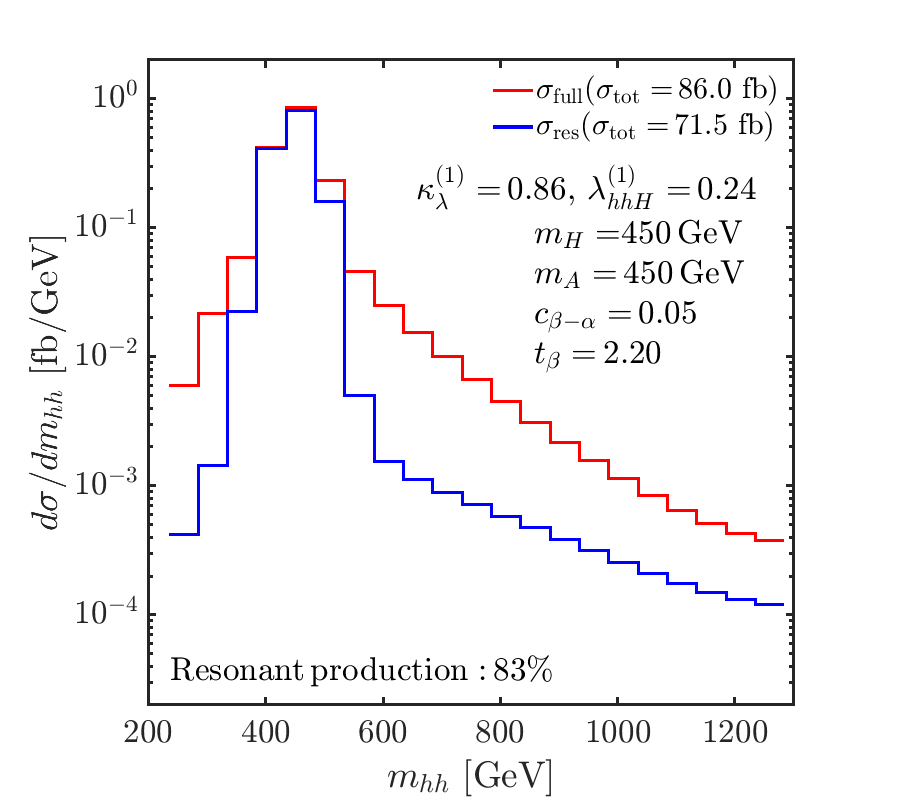}
	\end{center}
\caption{Same as Fig.~\ref{fig:bp1}, but for BP2 (allowed by non-resonant searches, excluded by resonant searches).}
\label{fig:bp2}
\end{figure}

While the benchmark point that we have discussed in \reffi{fig:mhhlimits} 
is unexcluded by the non-resonant and resonant searches, we now turn to two benchmark points that are claimed to be excluded by the existing resonant searches.    
In \reffis{fig:bp1} and \ref{fig:bp2} we show the results for the two benchmark points  
in the 2HDM type~I.  
For each case
we compare the \mhh\ distributions based purely on the resonant diagram, shown in blue, with the one based on the full
calculation, shown in red. In the displayed results the NLO results for the THCs have been used (with the values given in the 
respective plots). Like in the previous plots, all results are shown at LO in QCD. 
By comparing the predicted distributions based on the full result with the ones based on only the pure resonant contribution we will investigate to what extent the assumption of taking into account only the pure resonant contributions is justified.

\begin{table}[ht!]
\begin{center}
\begin{tabular}{c||c|c|c|c|c}
 & $t_\beta$ & $\CBA\,(\SBA>0)$  & $\MH$ & $\MA = \MHp$ & $\msq$ \tabularnewline\hline
BP1  & 4.0 & 0.05 & 450 & 800 & $\msq = \sfrac{\MH^2 \CA^2}{\TB}$ \tabularnewline
BP2  & 2.2 & 0.05 & 450 & 450  &  $\msq = \sfrac{\MH^2 \CA^2}{\TB}$ \tabularnewline
\end{tabular}
\caption{Input parameters for the selected benchmark points. Masses are given in GeV.
}
\label{tab:bp_params}
\end{center}
\end{table}

The input parameters for the selected benchmark points are defined in \refta{tab:bp_params}. These points 
have been obtained as
part of a broader scan of the 2HDM parameter space, on which we will report in more detail in a forthcoming publication.
We also give the total cross section values calculated with \texttt{HPAIR}
for the 
two benchmark points in \refta{tab:bpoints}. 
In column~2 and~3 we show the results of the full calculation at LO and NLO QCD, respectively (confirming the factor
of about~2 between them, as mentioned above). In column~4 and~5 we give the corresponding results taking into account only the resonant diagram. The 
cross section values at LO QCD quoted in the legends of the figures correspond to the integrated curves of \reffi{fig:bp1} and \ref{fig:bp2}.
Column~6 shows the ``obs.\ ratio", calculated with \texttt{HiggsTools} \cite{Bahl:2022igd,Bechtle:2008jh,Bechtle:2011sb,Bechtle:2013wla,Bechtle:2015pma,Bechtle:2020pkv,Bechtle:2013xfa,Bechtle:2014ewa,Bechtle:2020uwn}. 
The obs.\ ratio is defined as 
\begin{eqnarray}
\mbox{obs.\ ratio} \equiv \frac{\sigma^{\rm model}(ggH) \times {\rm BR}^{\rm model}(H\to hh)}{\sigma^{\rm obs}(ggH)\times {\rm BR}^{\rm obs}(H\to hh)} \;, 
\label{eq:obsdef}
\end{eqnarray}
where the 
superscript ``${\rm obs.}$" refers to the observed experimental limit and ``model" refers to the 2HDM. 
Here, the model cross sections have been calculated at NLO QCD in the Born improved heavy-top limit, 
using {\tt HPAIR}. The model branching ratios have been obtained with {\tt HDECAY} \cite{Djouadi:1997yw,Djouadi:2018xqq}, which we modified to include the effective NLO coupling $\lambda_{hhH} ^{(1)}$ in the decay width of the heavy Higgs boson into the SM-like Higgs boson pair. These calculated 2HDM cross section and branching ratio values are then provided as inputs for \texttt{HiggsTools}. 
The definition Eq.~(\ref{eq:obsdef}) implies that the points with an observed ratio larger than 1 are excluded by experimental searches. In view of the assumptions made in the experimental analyses we apply this limit only to the resonant contribution $\sigma^{\rm res}$. 
The benchmark points BP1 (\reffi{fig:bp1}) and BP2 (\reffi{fig:bp2}) are both excluded by the resonant search $pp \to hh \to b\bar{b}\tau^+\tau^-$ \cite{ATLAS:2022xzm}\footnote{This search is included in {\tt HiggsTools} dataset since version {\tt v1.6}.}.
\begin{table}[ht!]
\begin{center}
\begin{tabular}{c||c|c||c|c||c}
 & $\sigma {\rm \,(LO\, QCD)}$ & $\sigma {\rm \,(NLO\, QCD)}$ & $\sigma^{\rm res}  {\rm \,(LO\, QCD)}$ & $\sigma^{\rm res} {\rm \,(NLO\, QCD)}$ & ${\rm obs.\,ratio}$ \tabularnewline\hline
BP1  &  82.53 & 165.89 & 45.06 & 89.23 & 1.8 \tabularnewline
BP2  & 85.95 & 169.03 & 71.51 & 140.77 & 2.9\tabularnewline
\end{tabular}
\caption{Higgs pair production cross sections $\sigma (gg \to hh)$  [fb] and the resonant contribution only ($\sigma^{\rm res}$), computed with \texttt{HPAIR} at LO and NLO QCD in the Born improved heavy-top limit for the total cross section, respectively;  
``obs. ratio" obtained with \texttt{HiggsTools} (see text). 
}
\label{tab:bpoints}
\end{center}
\end{table}

Figure \ref{fig:bp1} shows the result for the benchmark point~BP1, which is
claimed to be excluded by resonant 
di-Higgs searches, 
but not by non-resonant searches. 
This point is characterized by 
significant corrections to 
\kala, corresponding to a parameter region
where the one-loop effective coupling approximation is well justified. Specifically, we find
\begin{eqnarray}
\kala\tree = 0.94, \; \kala\one = 5.01, \; \lahhH^{(0)} = 0.21 \mbox{ and } \lahhH\one = 0.23 \;,
\end{eqnarray}
\KR{
and
\begin{eqnarray}
\lambda_{hHH}^{(0)} = 0.37, \; \lambda_{HHH}^{(0)} = -0.06, \; \lambda_{hAA/hH^{\pm}H^{\mp}}^{(0)} = 7.56 \mbox{ and } \lambda_{HAA/HH^{\pm}H^{\mp}}^{(0)} = 0.30.
\label{eq:thcbpex}
\end{eqnarray}
}

The results given for the total cross sections 
indicate that the pure resonant contribution
amounts to about half of the full result (both at LO and NLO QCD), this is indicated with the percentage of the resonant production contribution in the full process, displayed in the bottom of \reffi{fig:bp1}. 
Concerning the \mhh\ distributions, one
can see that the qualitative features are similar to the right plot of 
\reffi{fig:mhhlimits}. While the pure resonant contribution shows a pronounced peak, this peak-like structure appears only as a rather small modulation of a smoothly falling distribution in the full result. As in \reffi{fig:mhhlimits} the cross section just above the $hh$ threshold is enhanced by several orders of magnitude compared to the expectation based on the pure resonant contribution. The peak-like structure in the full result will clearly be much more difficult to resolve experimentally than it would seem to be the case based on the pure resonant contribution. We therefore conclude that the exclusion limits obtained for the resonant di-Higgs searches by ATLAS and CMS may be too optimistic in view of the modifications that occur in the invariant \mhh\ mass distribution upon the inclusion of 
\GW{the SM-like non-resonant contributions that are present in all realistic scenarios and of the
relevant interference contributions}.

Our second example, BP2, is shown in \reffi{fig:bp2}, 
and defined by the input values in the second row of \refta{tab:bp_params}.
As BP1, it is claimed to be excluded by resonant di-Higgs searches, but not by the non-resonant ones.
Contrary to BP1, the higher-order corrections to the THCs are substantially smaller. We find 
\begin{eqnarray}
\kala\tree = 0.96, \; \kala\one = 0.86, \; \lahhH^{(0)} = 0.20 \mbox{ and } \lahhH\one = 0.24 \;,
\end{eqnarray}
\KR{
and
\begin{eqnarray}
\lambda_{hHH}^{(0)} = 0.15, \; \lambda_{HHH}^{(0)} = -0.05, \; \lambda_{hAA/hH^{\pm}H^{\mp}}^{(0)} = 0.12 \mbox{ and } \lambda_{HAA/HH^{\pm}H^{\mp}}^{(0)} = -0.05.
\label{eq:thcbpex}
\end{eqnarray}
}
For this parameter point the \mhh\ distribution based on the pure resonant contribution and on the full result
are more similar than in the previous example, and 
the pure resonant contribution amounts to about 83\% of
the full cross section. However, still a substantial broadening of the peak by the inclusion 
of the non-resonant diagrams can be observed. 
Similarly to BP1, we therefore conclude that the exclusion limits obtained for the resonant di-Higgs searches by ATLAS
and CMS are possibly too optimistic in view of the \mhh\ modifications due to the inclusion of all the relevant 
contributions in a realistic scenario.

\smallskip
Our discussion shows that the sensitivity of the resonant di-Higgs searches by ATLAS and CMS has already reached a level of sensitivity that strongly motivates to go beyond the assumption of restricting to the pure resonant contribution in deriving the experimental limits. A dedicated joint effort of experiment and theory would be desirable to define an appropriate framework in which the experimental limits should be presented in the future. \KR{In particular, the non-resonant contributions should be included in the signal model, and the possibility of interference effects between the resonant and non-resonant contributions should be incorporated. This will require an extension of the analysis setup involving additional parameters.}


\section{Conclusions}
\label{sec:conclusions}

The determination of the trilinear Higgs self-coupling as a first step towards elucidating the shape of the Higgs potential will be a prime goal of particle physics at the LHC and beyond. The current bounds on the trilinear Higgs self-coupling leave significant room for deviations of this coupling from the SM value. Such deviations in $\kala$ can occur in simple extensions of the SM such as the 2HDM, where they arise in particular from loop corrections involving additional Higgs bosons. While we have used the 2HDM as theoretical framework in our analysis, our qualitative results are applicable to a wide class of models of extended Higgs sectors.

In our analysis we have emphasized the need to compare the experimental results for di-Higgs production with precise theoretical predictions, in particular  including electroweak corrections besides QCD corrections, as they may lead to large effects in models with extended scalar sectors. Starting with an investigation 
of the experimental bounds that have been obtained from non-resonant di-Higgs production, we have investigated the impact of the loop contributions to $\kala$. Our results underline that, once the radiative corrections to the Higgs 
self-interactions are taken into account, the experimental bounds 
from the search for di-Higgs production at the LHC 
already provide sensitivity to parameter regions of the 2HDM that were 
unconstrained so far based on all other
existing experimental and theoretical limits.

We have then analyzed in detail the case where the di-Higgs production process receives a contribution from the resonant production of an additional neutral Higgs boson. The limits from those resonant searches that have been presented by ATLAS and CMS so far were obtained assuming a signal model consisting only of the resonant contribution, while the non-resonant SM-like contributions involving the $s$-channel exchange of the detected Higgs boson at 125~GeV as well as the box-type top-quark loop contribution have been neglected. Accordingly, the limits obtained by ATLAS and CMS can only be directly applied to scenarios where the impact of the non-resonant contributions 
is negligible compared to the pure resonant contribution. Using the 2HDM as a test case we have compared the full result for the \mhh\ invariant mass distribution, consisting of both the resonant and the non-resonant contributions as well as the interference effects and taking into account the loop corrections to the trilinear Higgs self-couplings $\lahhh$ and $\lahhH$, with the pure resonant contribution as used by ATLAS and CMS. In order to take into account the limited detector resolution in the experimental analyses we have applied a smearing 
of 15\% and a binning in \mhh\ of $50 \gev$ in our phenomenological study.

While the assumption of restricting to the pure resonant contribution made by ATLAS and CMS implies that the \mhh\ distribution corresponding to the assumed signal has a peak structure located at $\mhh \approx \MH$, the non-resonant contributions and the interference effects can modify this behavior. Indeed, we have found that the distributions based on the prediction arising from the full result can be significantly distorted as compared to the distribution that would be expected from the pure resonant contribution.  
Instead of a pronounced peak as it would be expected from the pure resonant contribution, we have demonstrated that the full result incorporating all relevant contributions can give rise to an \mhh\ distribution that is overall smoothly falling with just a small modulation near $\mhh \approx \MH$. The task to experimentally resolve 
this structure is clearly much more difficult than it would be the case if the distribution had the form as expected from the pure resonant contribution. We have pointed out the importance of the loop contributions to the trilinear Higgs self-couplings in this context. A striking feature related to the loop corrections to $\kala$ is a large effect on the differential cross section just above the $hh$ threshold. It arises because a large cancellation between the triangle and box form factors of the continuum diagrams that is present at the \mhh\ threshold at leading order no longer occurs upon the inclusion of the loop corrections.

In our numerical analysis we have specifically investigated examples of parameter points that would be classified as excluded according to the existing resonant searches and assessed to what extent the assumption of neglecting the non-resonant contributions made in obtaining the experimental limits is justified. Also in these cases we have found significant distortions of the distributions compared to the expectation from the pure resonant contribution. This implies that the exclusion limits obtained for the resonant di-Higgs searches by ATLAS and CMS may be too optimistic in view of the modifications that occur in the invariant \mhh\ mass distribution in realistic scenarios upon the inclusion of all the relevant contributions.

\smallskip
The results obtained in our paper indicate that the resonant di-Higgs searches carried out by ATLAS and CMS have meanwhile reached a level of sensitivity that strongly motivates to define an appropriate framework in which the experimental limits should be presented in the future. Avoiding the assumption of restricting to the pure resonant contribution in deriving the experimental limits, such a framework should make it possible to directly compare the experimental results with theoretical predictions in extended Higgs sectors. A dedicated joint effort of experiment and theory would seem to be desirable in this context.


\subsection*{Acknowledgements}

We thank Francisco Arco, Johannes Braathen and Michael Spira for useful discussions. The work of S.H.\ has received financial support from the
grant PID2019-110058GB-C21 funded by
MCIN/AEI/10.13039/501100011033 and by ``ERDF A way of making Europe", 
and in part by the grant IFT Centro de Excelencia Severo Ochoa CEX2020-001007-S
funded by MCIN/AEI/10.13039/501100011033. 
S.H.\ also acknowledges support from Grant PID2022-142545NB-C21 funded by
MCIN/AEI/10.13039/501100011033/ FEDER, UE.
The work of M.M.\ has been supported by the BMBF-Project 05H21VKCCA.
K.R.~and G.W.~acknowledge support by the Deutsche Forschungsgemeinschaft
(DFG, German Research Foundation) under
Germany's Excellence Strategy -- EXC 2121 ``Quantum Universe'' --
390833306. This work has been partially funded by the Deutsche
Forschungsgemeinschaft (DFG, German Research Foundation) - 491245950.

\normalem
\bibliographystyle{JHEP}
\bibliography{lit}

\end{document}